\documentclass[12pt]{iopart}

\usepackage{graphicx} 
\usepackage{amssymb}
\begin{document}

\title{Kinetic small angle neutron scattering of the Skyrmion lattice in MnSi}

\author{S. M\"uhlbauer$^1$, J. Kindervater$^2$, T. Adams$^2$, A. Bauer$^2$, U. Keiderling$^3$, 
and C. Pfleiderer$^2$}

\address{$^1$ Heinz Maier-Leibnitz Zentrum (MLZ), Technische Universit\"at M\"unchen, D-85748 Garching, Germany}
\address{$^2$ Physik-Department, Technische Universit\"at M\"unchen, D-85748 Garching, Germany}
\address{$^3$ Helmholtz Zentrum Berlin, BENSC, D-14109 Berlin, Germany}
\ead{Sebastian.muehlbauer@frm2.tum.de}
\vspace{10pt}
\begin{indented}
\item[]February 2014
\end{indented}

\begin{abstract} We report a kinetic small angle neutron scattering study of the skyrmion lattice (SL) in MnSi. Induced by an oscillatory tilting of the magnetic field direction, the elasticity and relaxation of the SL along the magnetic field direction have been measured with microsecond resolution. For the excitation frequency of $325\,{\rm Hz}$ the SL begins to track the tilting motion of the applied magnetic field under tilting angles exceeding $\alpha_c\gtrsim0.4^{\circ}$. Empirically the associated angular velocity of the tilting connects quantitatively with the critical charge carrier velocity of $\sim 0.1\,{\rm mm/s}$ under current driven spin transfer torques, for which the SL unpins. In addition, a pronounced temperature dependence of the skyrmion motion is attributed to the variation of the skyrmion stiffness. Taken together our study highlights the power of kinetic small angle neutron scattering as a new experimental tool to explore, in a rather general manner, the elasticity and impurity pinning of magnetic textures across a wide parameter space without parasitic signal interferences due to ohmic heating or Oersted magnetic fields.
\end{abstract}

%
%
%
%
%

\section{Introduction}

The observation of an unusual rotation of the neutron diffraction pattern of the so-called A-phase in MnSi under an electric current applied perpendicular to the magnetic field stabilising the A-phase \cite{Jonietz:08,Jonietz:10}, resulted in the first identification of a skyrmion lattice as a new form of magnetic order \cite{muehlbauer:09b,Muehlbauer:diss}. Representing a kind of magnetic whirls, skyrmions are characterised by a non-vanishing topological winding number as the new aspect. Whereas the existence of skyrmion lattices had been anticipated in a micro-magnetic mean-field model for chiral magnets with natural or superimposed uniaxial anisotropies \cite{Bogdanov:89}, the discovery of a spontaneous skyrmion lattice phase in a cubic compound with very weak magnetic anisotropies was unexpected \cite{Bogdanov:94}. 

A series of comprehensive studies have by now established beyond doubt the early conjecture, that the skyrmion lattice in cubic chiral magnets is generic and stabilised by the effects of thermal fluctuations \cite{Janosch:13, Bauer:13_PRL, Bauer:12_PRB, kindervater:14, Bauer:Spr, Buhrandt:13}. The magnetic properties are thereby the result of a set of hierarchical energy scales with ferromagnetic exchange coupling on the strongest scale, isotropic Dzyaloshinsky-Moriya spin-orbit interactions on intermediate scales and higher order spin orbit coupling terms on the weakest scale \cite{Landau-Lifshitz:vol8}. Careful measurements avoiding inhomogeneities of the applied magnetic field due to the effects of demagnetisation reveal, that the skyrmion lattice phase is ordered exceptionally well with resolution-limited magnetic coherence length exceeding at least several $10^{-6}\,{\rm m}$ \cite{Adams:11}.

Following another two years of detailed measurements the original observation of a rotation of the diffraction pattern under electric current flow could finally be attributed unambiguously to the combination of (i) an exceptionally efficient coupling originating in the spin of the charge carriers with the non-trivial topology of the spin order, (ii) very weak pinning by defects, and (iii) a small temperature gradient in the sample generated by ohmic heating \cite{Jonietz:10,Jonietz:Diss}. The latter causes a gradient in the spin currents across the SL domains, generating a torque that is balanced by the very weak magnetic anisotropies when the SL unpins and moves with respect to the crystal structure. The complexity of the interplay between the conduction electrons and SL may be elegantly accounted for in the language of an emergent electrodynamics, where the presence of each skyrmion is viewed in terms of a fictitious magnetic field of precisely one flux quantum ($h/\rm{e}$). This fictitious magnetic field gives rise to a topological Hall effect \cite{Neubauer:09,Franz:14}, as well as an emergent electric field when the SL moves \cite{Schulz:12}. 

The first identification of a skyrmion lattice as a new form of magnetic order and, perhaps more importantly, the discovery of spin torque effects at ultra-low current densities, have motivated a wide range of experimental and theoretical studies \cite{2013:Nagaosa:NN}.  Yet, an important  unresolved question concerns the details of the unpinning of the skyrmion lattice at the critical current density $j_c$. Further, recent theoretical work suggests, that the unpinned skyrmion lattice for current densities $j>j_c$ displays enhanced long-range order and thus reduced mosaicity as compared with the pinned skyrmion lattice \cite{Reichardt:15}. However, an inherent difficulty in systematic studies of the pinning mechanisms under spin torques is the presence of ohmic heating and stray magnetic fields due to the applied currents, which, in addition, vary strongly as a function of current density.  Moreover, studies of the interaction of spin currents with skyrmion lattices as generated by charge currents are not possible at all in electrical insulators. Such a situation has recently been discussed for the skyrmion lattice in Cu$_2$OSeO$_3$, where the observation of a rotational motion observed in Lorentz-force TEM data was attributed to a current of magnons \cite{2014:Mochizuki:NatMat}.  

In this paper we report proof-of-concept kinetic small angle neutron scattering measurements in MnSi, the most extensively studied skyrmion lattice material to date. The experimental method we report is based on a skyrmion lattice motion generated by means of periodically oscillating the magnetic field direction, which stabilises the skyrmion lattice. The same concept was previously used for studies of the elastic tilt modulus $c_{44}$ of the superconducting vortex lattice in ultra-pure Nb \cite{Muehlbauer:11_PRB}. As its main advantage, the oscillation of the field direction drives a skyrmion lattice motion without need for an electric current to generate spin currents. Thus studies of the unpinning and onset of the skyrmion lattice motion become possible even in electrical insulators. 

However, unlike the translational motion of a skyrmion lattice generated by the spin torques of a spin current of spin-polarised conduction electrons or magnon flow, the oscillatory tilting of the field direction yields a distribution of lateral skyrmion lattice displacements across the size of typical skyrmion lattice domains. A complete analysis requires therefore, in principle, the full deconvolution of the scattering data by these distributions, which is clearly beyond the work reported in the following. Nonetheless the preliminary results provide already deep insights, that highlight the potential of kinetic small angle scattering in the exploration of defect-related pinning of magnetic order, notably skyrmion lattices and related textures.

\section{Experimental Methods}
Our kinetic small angle neutron scattering (SANS) studies were performed at the beam-line V4 at HZB, Berlin. For the measurements of the Skyrmion lattice in MnSi the neutron beam of V4 was collimated over a distance of 12\,m. The detector was placed at a distance of 8\,m. A wavelength of 5.92\,{\AA} was chosen with a wavelength spread of 10\%. In addition to the standard SANS setup the instrument V4 may be operated for so-called 'TIme resolved SANS Experiments' (TISANE), offering a time resolution of a few microseconds. A detailed description of the TISANE technique may be found in Ref.\,\cite{Wiedenmann:06}. In TISANE mode the beam is pulsed by a dedicated multi-window chopper placed close to the instruments velocity selector. TISANE is a stroboscopic neutron scattering technique, where the sample property of interest is oscillated by an external stimulus. The neutron counts are recorded in event mode lists, and the signal integrated over many cycles of the stimulus. Due to the pulsed beam structure the wavelength smearing of the time-resolved signal can be corrected and time-binned posterior to the measurements. The resulting signal hence is averaged over many cycles of the stimulus. For our experiment the amplitude of an oscillating magnetic field served as the stimulus. 

\begin{figure}
\begin{center}
\includegraphics[width=0.85\textwidth]{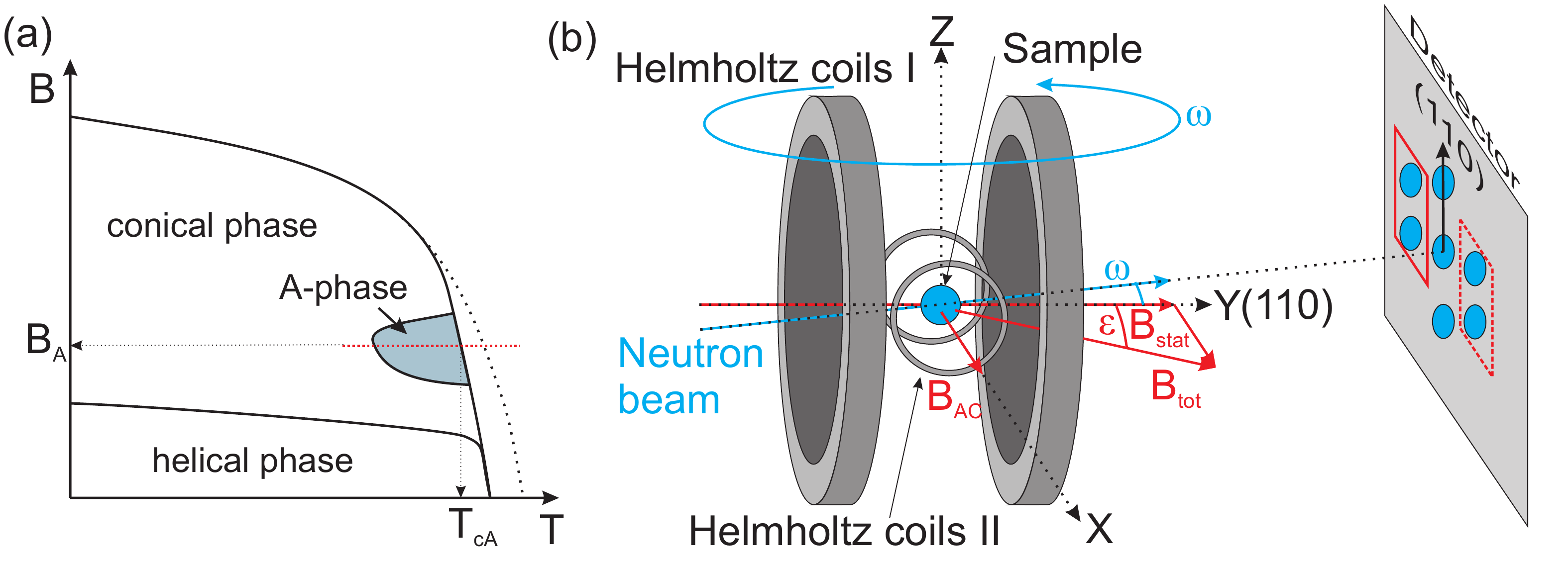}
\caption{(a) Schematic depiction of the magnetic field / temperature phase diagram of MnSi. The dashed horizontal red line indicates the temperature scan across the Skyrmion lattice phase at the magnetic field $B_A$, where time resolved kinetic measurements have been performed. The transition from the Skyrmion lattice phase to the paramagnetic phase at $B=B_A$ is denoted $T_{cA}$. (b) Illustration of the experimental setup with the orientation of static and AC magnetic field: The static magnetic field $B_{stat}$ was aligned almost parallel to the incoming neutron beam, tilted with respect to the vertical rocking axis by $\omega$. The oscillating magnetic field was generated by a small set of Helmholtz coils inside the main coil. The AC field $B_{AC}$ was aligned perpendicular to $B_{stat}$. As $B_{stat} \gg B_{AC}$ the resulting total field $B_{tot}$ may be described by a small oscillation of $B_{stat}$ by the angle $\epsilon$ around its equilibrium position. The spherical sample was placed in the center of both Helmholtz coils with an (110) direction parallel to the incoming neutron beam and an (110) axis vertical. The six-fold scattering pattern of the skyrmion lattice is depicted schematically on the detector assuming a suitable small angle neutron scattering geometry.}
\label{Fig_1}
\end{center}
\end{figure}

The sample was cooled with a closed cycle cryostat equipped with a quartz vacuum shield  reaching a base temperature of 10\,K. A quartz tube was used to eliminate shielding of the AC field, normally expected for cryostat windows made of Al as used widely. A second,  radiation shield made of Al was removed for similar reasons. A schematic depiction of the experimental setup is shown in Fig.\,\,\ref{Fig_1}. The static main magnetic field $B_{stat}$, generated by water cooled copper coils was aligned approximately parallel to the incoming neutron beam. Driven by a frequency generator and AC-amplifier a second set of small Helmholtz coils was placed inside the main coil, generating a sinusoidally oscillating magnetic field of a few mT perpendicular to $B_{stat}$. With a main field $B_{stat}=$183\,mT and a oscillating field component of a few mT perpendicular to the main field, the superposition of both fields corresponds to a small oscillation of the magnetic field direction by an angle $\epsilon$ with respect to the equilibrium position. The entire setup including cryostat and both coils could be rotated with respect to the vertical axis by the rocking angle $\omega$.

As mentioned above the frequency dependence and shielding effects of the oscillating field component were tracked in terms of the voltage $V_{pp}$ induced in a set of small pick-up coils, placed both inside and outside the cryostat. Moreover, small frequency dependent AC-heating effects (less than 0.05\,K) caused by the AC field were observed. The temperature scale was calibrated by the loss of intensity associated with the transition from the A-phase to the paramagnetic phase at $T_{cA}$ for each frequency and amplitude of the AC-field. Temperatures are hence given on a relative scale ($T_s=T-T_{cA}$).

A high quality MnSi single crystal sample with a spherical shape (diameter 5\,mm) was used for our measurements. The spherical sample shape ensured homogeneous, parallel demagnetizing fields across the entire sample volume under all angles $\epsilon$ of the applied magnetic field. The single crystal MnSi sample was prepared by optical float-zoning under ultra-high vacuum compatible conditions \cite{Neubauer:RSI}, resulting in the usual very high sample purity. A (110) crystallographic direction was aligned parallel to the static component of the magnetic field $B_{stat}$. The sample was aligned additionally such that a further (110) direction coincided with the vertical axis.

\section{Experimental Results}

Data were recorded in the static mode and the in time-resolved TISANE mode driven by the oscillating magnetic field orientation. Before we turn to the time resolved data, it proves to be helpful to summarise data recorded in the standard SANS mode without oscillating field component. 

\begin{figure}
\begin{center}
\includegraphics[width=0.85\textwidth]{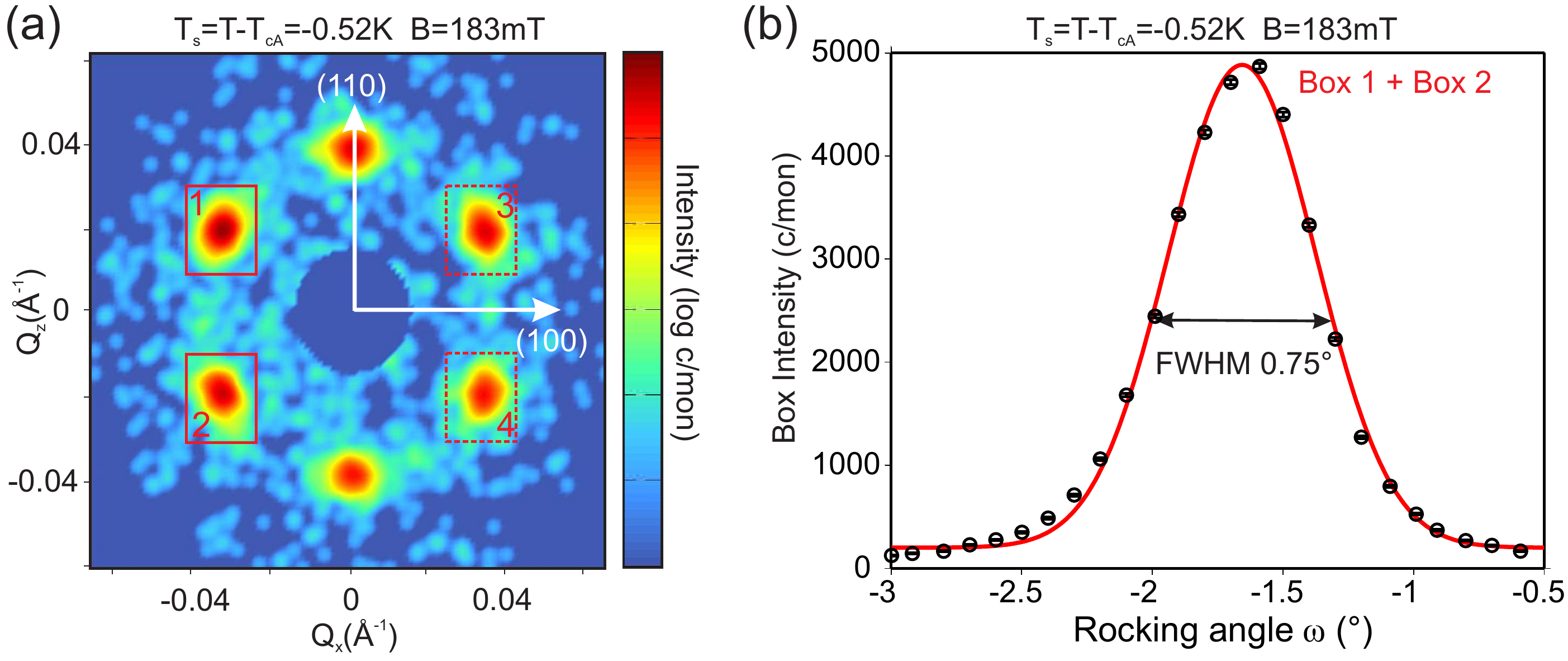}
\caption{Panel(a): SANS scattering pattern of the Skyrmion lattice of MnSi, recorded at a temperature of $T_s=-0.52$\,K and a static magnetic field of 183\,mT. Magnetic field and incoming neutron beam were parallel and aligned along a crystallographic (110) direction, an additional (110) axis was aligned vertically. The intensity is color-coded on a logarithmic scale. No oscillation of the magnetic field was applied. The red boxes labelled 1 through 4 indicate the area of integration for rocking scans around the vertical axis ($\omega$). A typical rocking scan of box 1 and box 2 is shown in panel (b). The red line corresponds to a fit to a Gaussian function with a FWHM of 0.75$^{\circ}$ is observed.}
\label{Fig_2}
\end{center}
\end{figure}

Fig.\,\ref{Fig_2} summarises typical SANS data, recorded for a magnetic field $B_{stat}=$183\,mT and a temperature $T_s=$-0.52\,K well inside the A-phase. The characteristic six-fold scattering pattern of the Skyrmion lattice in MnSi may be readily seen in Fig.\,\ref{Fig_2}\,(a). For a (110) crystallographic direction along the magnetic field, the symmetry axis of the hexagonal scattering pattern is also a (110) direction within the plane perpendicular to the magnetic field \cite{Adams:11}. Fig.\,\ref{Fig_2}\,(b) shows a typical rocking scan with respect to $\omega$. The area of integration is indicated by the red boxes sketched around the left/right pairs of peaks in Fig.\,\ref{Fig_2}\,(a). The rocking curve is well described by a Gaussian shape with a rocking width of FWHM 0.75$^{\circ}$. This corresponds to the instrumental resolution and corresponds to a well-ordered Skyrmion lattice as reported in the literature \cite{Adams:11}. We note, that  exponential rocking scans were observed in cylindrical samples due to the inhomogeneities of the demagnetising fields as reported elsewhere \cite{muehlbauer:09b,Adams:11}.

\begin{figure}
\begin{center}
\includegraphics[width=0.95\textwidth]{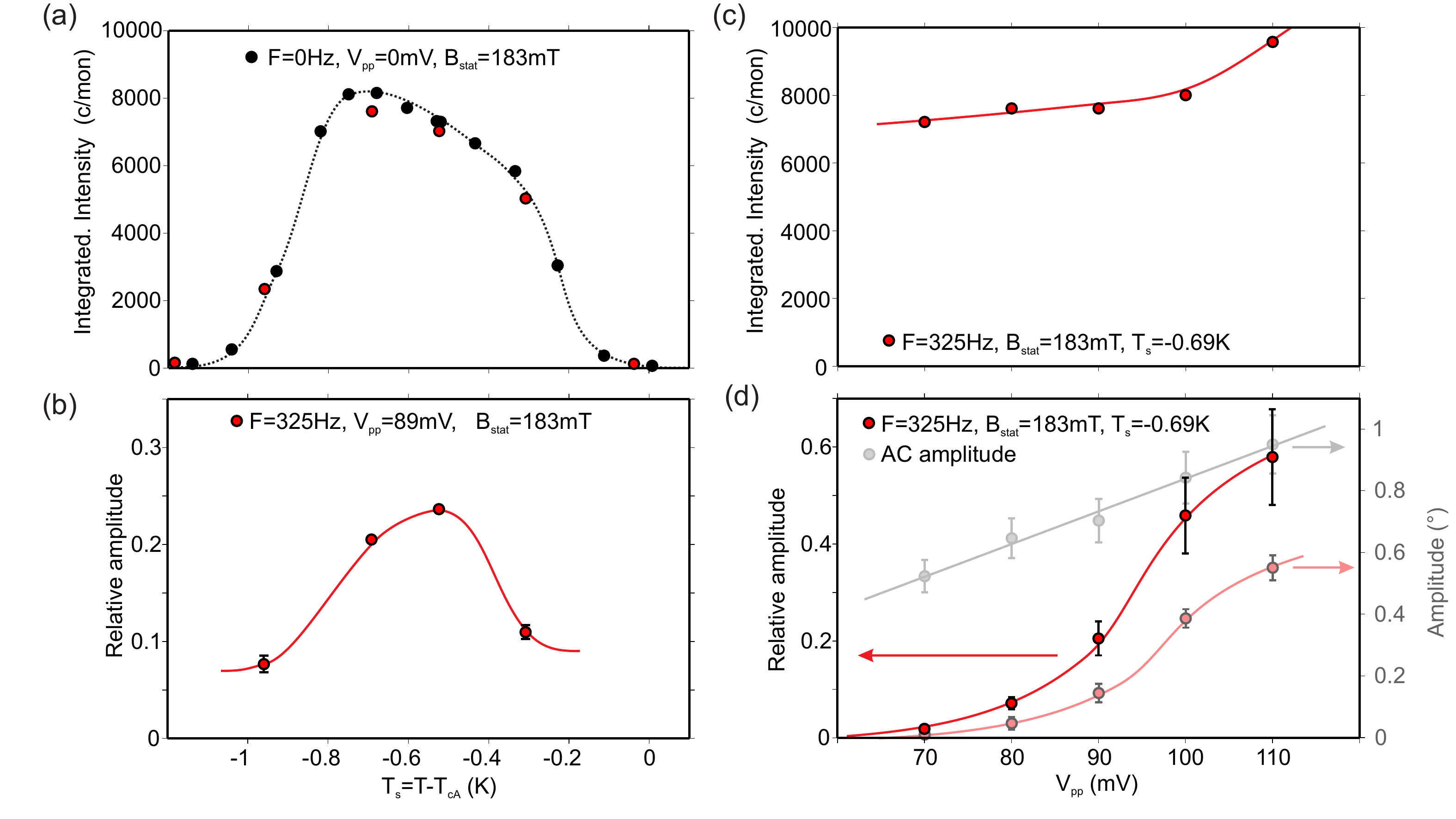}
\caption{Panel (a): Temperature dependence of the integrated intensity of the Skyrmion lattice Bragg peaks for $B_{stat}=$183\,mT for absent field oscillation and for a frequency of 325\,Hz. The temperature scale is adjusted to the static data to account for the eddy current heating effects caused by the AC field. Panel (b) depicts the corresponding oscillation amplitude of the Skyrmion lattice for a frequency of 325\,Hz. Panel (c) and (d) show the integrated intensity and the oscillation amplitude for a fixed frequency of 325\,Hz and various AC amplitudes.}
\label{Fig_5}
\end{center}
\end{figure}

The temperature dependence of the integrated intensity of the diffraction peaks of the Skyrmion lattice will be addressed further below, in Fig.\,\ref{Fig_5}\,(a) together with time resolved data, again for $B_a=183\,{\rm mT}$. The data has been obtained on warming the sample after field cooling in $B_a=183\,{\rm mT}$.

We now turn to the results of our time resolved kinetic experiment. Data were recorded as a function of temperature for a static field of $B_a=183\,{\rm mT}$ and an AC field perpendicular to this static field with an amplitude between 3 and 6\,mT driven at a frequency of 325\,Hz. 

\begin{figure}
\begin{center}
\includegraphics[width=0.75\textwidth]{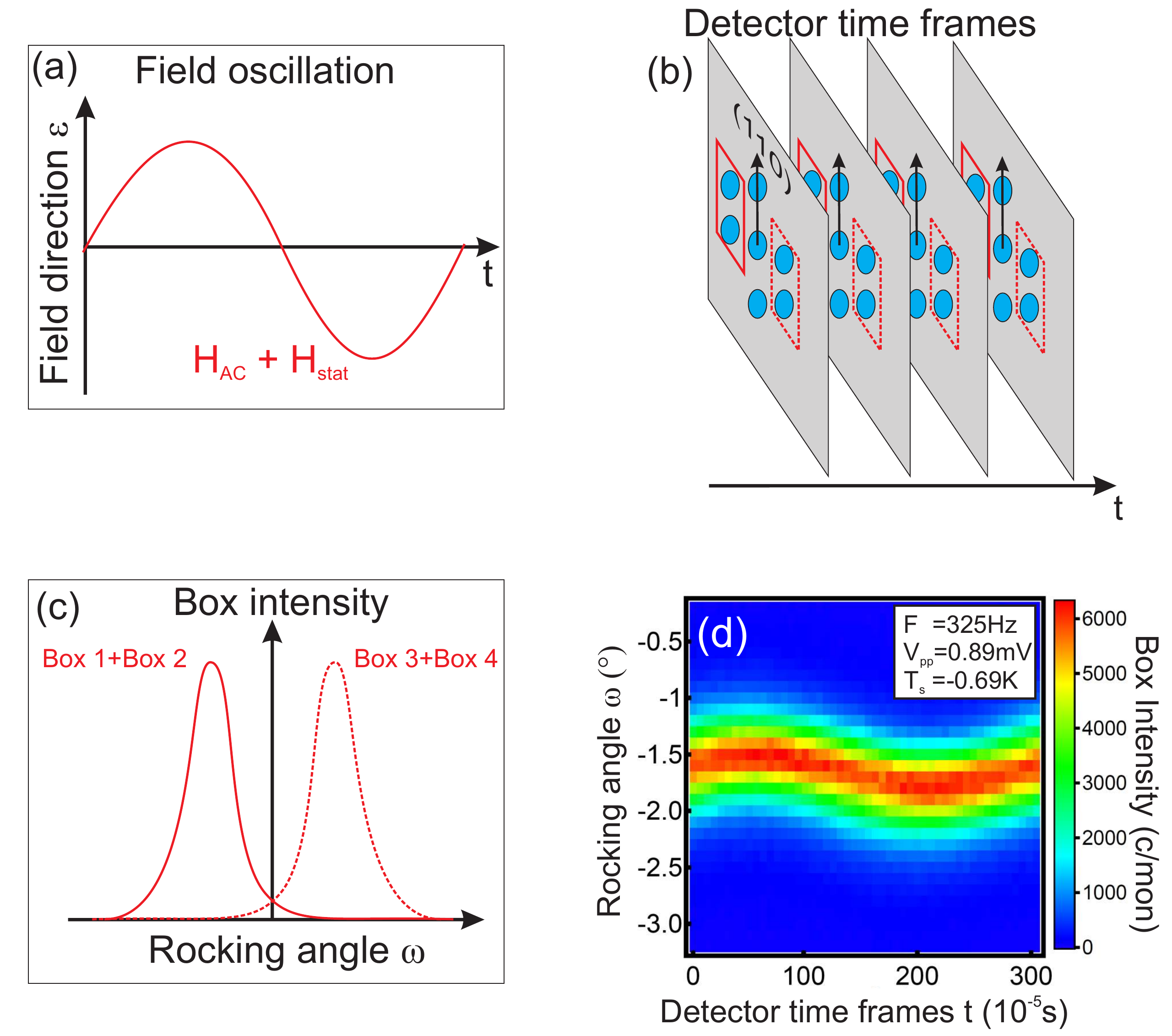}
\caption{Schematic depiction of data processing in a time resolved TISANE measurement. Panel (a) shows a single cycle of the oscillating field direction as obtained due to the addition of $B_{AC} + B_{stat}$. In a typical stroboscopic measurement, many of these cycles are added to collect enough neutron counts per time-bin. The different detector time-bins are illustrated in panel (b). The integration boxes on the left and right pair of diffraction peaks are indicated by the red boxes. A typical rocking scan of the skyrmion lattice with respect to the vertical axis $\omega$ is shown in panel (c): The intensity in both red integration boxes is depicted as function of rocking angle. Panel (d) finally depicts time resolved rocking maps of the Skyrmion lattice. The angular position of the Skyrmion lattice for the right integration box is plotted as a function of time for a cycle of the field oscillation. A vertical cut in the map corresponds to a rocking scan for a particular time-bin.}
\label{Fig_3}
\end{center}
\end{figure}

Before we discuss our results in detail, it is helpful to introduce the nomenclature used for our data analysis. Fig.\,\,\ref{Fig_3} schematically illustrates the data processing and data reduction: Fig.\,\ref{Fig_3}\,(a) shows a cycle of the oscillating magnetic field direction as a function of time. As typical for a stroboscopic measurement many such cycles are recorded to increase the statistics. Our data hence represent the statistical average over many field oscillations. The data is next processed in terms of 50 individual time bins, with a bin length of $\approx\,61.5\,{\rm \mu s}$ for a frequency of 325\,Hz. To avoid possible problems caused by a lack of frame overlap the first few cycles of each measurement were discarded. 

The intensity contained in the integration box over the right pair of peaks was evaluated as a function of time. The measurement was repeated accordingly for a set of rocking angles $\omega$ over a range of 3$^{\circ}$ with a step size of 0.2$^{\circ}$, essentially covering the set of angles of a regular rocking scan of a diffraction peak of the Skyrmion lattice. Such a schematic rocking scan is shown in Fig.\,\ref{Fig_3}\,(c). The two-dimensional intensity distribution shown in Fig.\,\ref{Fig_3}\,(d) hence reflects the angular position (rocking scan) of the Skyrmion lattice as a function of time over one cycle of the field oscillation averaged over many periods of the oscillating magnetic field. As the $k$-vectors of the Skyrmion lattice are oriented in a plane perpendicular to the magnetic field, our data shows how the Skyrmion lattice responds to a distortion of the field direction, hence probing the elastic stiffness and correlation of the Skyrmion lattice along the field direction. As a SANS-experiment is sensitive to the entire sample volume, our data reflect the integrated signal over the entire sample.

\begin{figure}
\begin{center}
\includegraphics[width=0.85\textwidth]{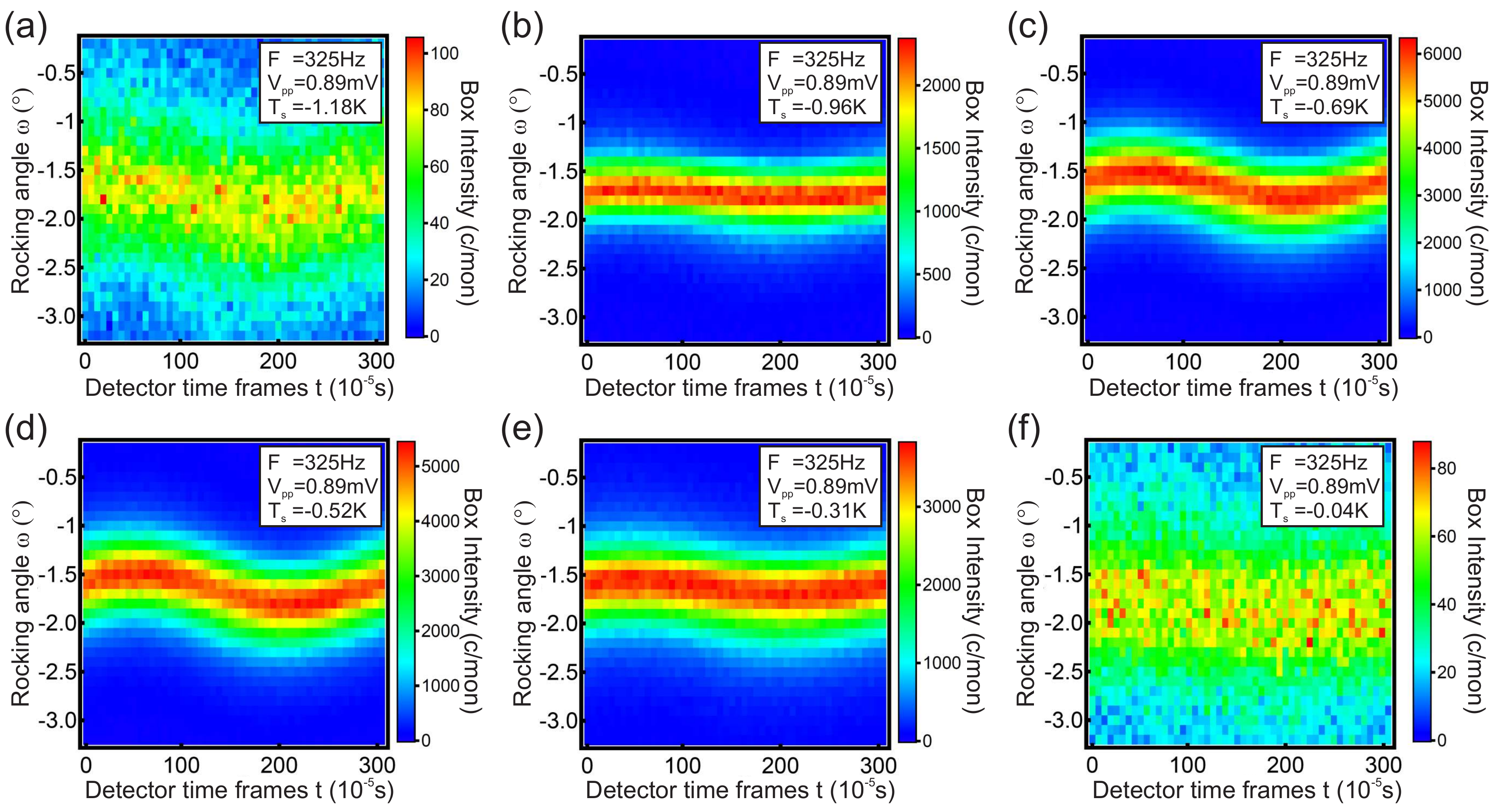}
\caption{Time resolved rocking maps for different temperatures across the A-phase of MnSi as stated in each panel. A static magnetic field of $B_{stat}=$183\,mT and and oscillating field component $B_{AC}=$3\,mT for a frequency of $F=$325\,Hz were applied.}
\label{Fig_4}
\end{center}
\end{figure}

Fig.\,\ref{Fig_4} summarises data recorded for increasing sample temperature across the A-phase of MnSi. An oscialltion frequency of 325\,Hz, a static field of 183\,mT and an amplitude of the oscillating field of 4.6\,mT were applied. Along the time axis 50 time bins each with a length of $\sim\,61.5\,{\rm \mu\,s}$ were used. The oscillation of the Skyrmion lattice is clearly visible, with increasing amplitude in the center of the A-phase. At the borders of the A-phase, the intensity vanishes and the mosaic spread increases. We attribute the latter to the phase coexistence of the skyrmion lattice with the conical phase at phase boundary at low temperatures and the paramagnetic state at high temperatures.

A quantitative analysis of the neutron scattering intensity in the Skyrmion lattice phase for increasing temperature across the A-phase and different frequencies of the AC-field are summarized in Fig.\,\ref{Fig_5}. Shown in Fig.\,\ref{Fig_5}\,(a) is a comparison of the integrated intensity of the Skyrmion diffraction peaks as recoded in kinetic measurements at 325\,Hz, as well as data recorded in the static mode. To correct a small systematic offset in the sample temperature due to eddy current heating the temperature axis was systematically shifted by 0.04\,K for 325\,Hz. Data are therefore shown on a relative temperature scale. Apart from this small systematic correction, the temperature dependence of the static and kinetic data are in excellent agreement with each other as well as the literature, where lines serve as a guide to the eye. 

To infer further information the oscillation of the Skyrmion lattice as observed by the time resolved rocking maps was fitted by a sinusoidal function. The amplitude of the skyrmion oscillation, given as an angle in degrees, was normalised by the amplitude $\epsilon$ of the change of orientation of the combined static and oscillating magnetic field in degrees. In the following this ratio is referred to as the \textit{relative amplitude}. 

Shown in Fig.\,\ref{Fig_5}\,(b) is the relative amplitude of the skyrmion lattice oscillation as a function of relative temperature. The temperature dependence of the relative oscillation amplitude of the Skyrmion lattice displays a clear clear maximum in the center of the A-phase, where lines serve as a guide to the eye. In contrast, at the borders of the A-phase, where the skyrmion lattice intensity decreases, the relative amplitude decreases substantially.  

Shown in Figs.\,\ref{Fig_5}\,(c) and (d) are the dependence of the integrated intensity as well as the amplitude and the relative amplitude on the AC-amplitude, respectively, where data were recorded for a constant temperature of $T_s=-0.67\,{\rm K}$. For systematic studies of the amplitude dependence of the skyrmion lattice response in the centre of the A-phase we have taken into account the effects of eddy current heating with increasing amplitude of the AC field, keeping the sample temperature constant. 

For small oscillation amplitudes of the magnetic field orientation the integrated intensity corresponds to the value found in the static data and the skyrmion lattice does not respond at all to changes of field orientation. Further, as shown in Fig.\,\ref{Fig_5}\,(c) the integrated intensity increases slightly with increasing oscillation amplitude up to the highest values studied. This compares with the amplitude of the skyrmion lattice reorientation as a function of oscillation amplitude, which is zero up to a critical angle $\epsilon_c\approx 0.4^{\circ}$. For amplitudes exceeding $\alpha_c$ the oscillation amplitude increases rapidly and appears to grow with a slope that scales with the increase of oscillation amplitude by a factor of order two. Hence the behaviour of the skyrmion lattice displays all the characteristics expected of an unpinning transition. As the integrated intensity even increases slightly across this unpinning transition, the difference of oscillation amplitude and skyrmion lattice amplitude is clearly not related to part of the skyrmion lattice still being pinned. 

\section{Discussion}

Two main results of our kinetic small angle neutron scattering studies of the skyrmion lattice in MnSi can be identified: First, we observe a pronounced temperature variation of the sykrmion motion for constant AC frequency and amplitude of the oscillatory field as seen in panel (b) of Fig.\,\,\ref{Fig_5}: Counterintuitively, the largest amplitude of the skyrmion motion is observed in the center of the A-phase with almost negligible movement for both increasing and decreasing temperature. However, in comparison to static data given in panel (a) the integrated intensity recorded at 325\,Hz unambigously proves that the skyrmion motion which was observed still represents bulk behaviour of the sample as a considerable fraction of the scattered intensity is left. The temperature variation seen in panel (b) hence cannot be tracked down simply to surface effects or phase coexistance. Moreover, as the frequency and the amplitude of the tilting magnetic field remain constant over the entire temperature scan and the resitivity of MnSi is not supposed to change significantly over the rather small temperature window of the A-phase, we strongly believe that the behaviour observed directly represents the temperature dependence of the skyrmion lattice stiffness along the magnetic field direction. Here, a clear maximum is expected at the center of the A-phase:

With inceasing skyrmion stiffness, a bending of the skyrmion lines is less benefitial. This leads to a larger motion of the skyrmion lattice in a two-fold manner: (i) As our neutron scattering experiment always integrates the entire sample, a uniform movement of a stiff skyrmion lattice will raise a larger signal as compared to a strong bending right at the surfece of the sample. (ii) Collective pinning of the skyrmion lattice is less efficient for a rigid skyrmion lattice, also leading to larger motion.

Second, clear evidence of an unpinning transition in the centre of the skyrmion lattice phase at an oscillation amplitude of $\epsilon_c\approx 0.4^{\circ}$ at 325\,Hz is identified. Phenomenologically the behaviour observed may be the result of several aspects, notably (i) the coupling between the skyrmion lattice orientation and the direction of the applied magnetic field, (ii) pinning the skyrmion lattice by defects, (iii) the strength and the character of the magnetic anisotropies, and (iv) spin currents that originate in the eddy currents induced by the AC field.

As the magnetic properties are due to a hierarchy of energy scales, the stabilisation of the skyrmion lattice perpendicular to the applied field is energetically dominating to leading order. In a recent study we have systematically mapped out the precise orientation and alignment of the skyrmion lattice as a function of field direction with respect to the crystal lattice. A combination of terms in the free energy that are fourth and sixth order in spin-orbit coupling account fully for the orientation of the skyrmion lattice within the plane perpendicular to the applied magnetic field as well as a tiny tilt of the skyrmion lattice plane away from being perfectly perpendicular to the applied magnetic field of no more than a few degrees \cite{Adams:diss,adams2016}. For the field direction studied in our kinetic SANS measurements, where $B_{stat}$ was parallel to $\langle110\rangle$ these tiny tilts vanish and the potential landscape is particularly flat. 

On these grounds the skyrmion lattice is expected to follow accurately even in the limit of vanishingly small changes of field direction. It is instructive to consider also the possibility of spin-transfer torques associated with the (electric) eddy currents induced by the AC field. Modelling the sample as a single-turn coil with a diameter of 5\,mm perpendicular to the AC field, assuming a diameter of the wire of 2\,mm, and taking into account the resistivity of MnSi around $30\,{\rm K}$ of approx 35 $\mu \Omega {\rm cm}$, the AC field induces electric current densities of order ${\rm A\,m^{-2}}$. This value is well below the critical current densities of $10^6\,{\rm A m^{-2}}$ above which spin transfer torque effects have been observed and we expect no effect of this kind. 

Thus the complete lack of a response for changes of field orientation below $\epsilon_c$ and the marked increase of the response above $\epsilon_c$ contrasts the expected close tracking between the skyrmion lattice orientation and the orientation of the applied field. Phenomenologically the behaviour is characteristic of an unpinning from defects. In fact, it is interesting to note that recent positron annihilation based spectroscopic studies establish point defect concentrations of our MnSi samples with a mean spacing of the defects below 10\,nm \cite{Reiner:PhD,Reiner:preprint}. Thus the spacing of the pinning sites is small as compared to the helical modulation length of $\sim20\,{\rm nm}$ and well below the typical size of skyrmion lattice domains. 

It is finally instructive to infer the minimum size $d$ of a skyrmion lattice domain, on the surface of which the skyrmion lattice moves laterally when tilting the field with the critical charge carrier velocity $v_c$ characteristic of current driven spin torques. This critical velocity corresponds roughly to $v_c\approx 0.1\,{\rm mm\,s^{-1}}$. Based on some simple estimates this is the case for skyrmion lattice domains exceeding $d\approx70\,{\rm \mu\,m}$. Even though this value of $d$ is quite large, it is nonetheless consistent with high-resolution SANS experiments \cite{Adams:PRL2011}.

\section{Conclusion}
In conclusion, we reported kinetic small angle neutron scattering of the skyrmion lattice pinning and elasticity in MnSi. As our main result we find clear evidence of a skyrmion lattice unpinning. Empirically the critical angular velocity of the tilting connects with the critical SL velocity of $\sim 0.1\,{\rm mm/s}$ observed under current driven spin transfer torques. Moreover, we observe a quite pronounced temperature variation of the skrmion motion for constant oscillatory field that tracks the skyrmion stiffness. Thus our study highlights the power of kinetic small angle neutron scattering as an experimental probe to explore, in a rather general sense, the pinning and elasticity of magnetic textures across a wide parameter space without parasitic signal interferences due to ohmic heating or Oersted magnetic fields.

\section{Achowledgments}
We wish to thank P. B\"oni, M. Garst and A. Rosch for fruitful discussions. Financial support through DFG TRR80 (From Electronic Correlations to Functionality) and ERC AdG (291079, TOPFIT) is gratefully acknowledged. J.K., T.A., and A.B. \ acknowledge financial support through the TUM graduate school.

\section*{References}
\bibliography{smuehlba_bib}

\end{document}